\documentclass[twocolumn]{article}
\begin{document}

\title{Topological Black Holes in Quantum Gravity}

\author{J.\ Kowalski--Glikman\thanks{e-mail: jurekk@ift.uni.wroc.pl}\  \
and
D.~Nowak--Szczepaniak\thanks{e-mail: dobno@ift.uni.wroc.pl}\\
Institute for Theoretical Physics\\ University of Wroc\l{}aw\\
Pl.\ Maxa Borna 9\\ Pl--50-204 Wroc\l{}aw, Poland}

\maketitle

\abstract{ We derive  the   black hole solutions with horizons of
non-trivial topology and investigate their properties in the
framework of an approach to quantum gravity being an extension of
Bohm's formulation of quantum mechanics. The solutions we found
tend asymptotically (for large $r$)  to  topological black holes.
We  also analyze the thermodynamics of these space-times.}

\section{Introduction}

In the recent paper \cite{Jurekbh} we started preliminary
investigations of black hole solutions of quantum gravity. In this
paper we would like to extend this investigations to the case of
the so--called topological black holes, being generalizations of
the standard Schwarzschild black hole solution of general
relativity, characterized by non-trivial topology of the horizons.
The constant time sections of such space-times are of the generic
form
\begin{equation}\label{1}
 ds^2_3= A(r) dr^2 + r^2 d^2\Omega,
\end{equation}
where $d^2\Omega$ is a metric of a Riemann surface of genus
$1,0,-1$, which in coordinates $\theta$, $\phi$ reads $$ d^2\Omega
=d\theta^2 + \sin^2\theta d\phi^2, \quad k=1, \quad
\mbox{sphere};$$ $$ d^2\Omega =  d\theta^2 +  d\phi^2, \quad k=0,
\quad \mbox{torus}; $$ $$ d^2\Omega =  d\theta^2 + \sinh^2\theta
d\phi^2, \quad k=-1, \quad \mbox{pseudo-sphere}.$$ Such
space-times and their thermodynamics have been analyzed recently
from many perspectives in papers \cite{aminneborg}, \cite{mann},
\cite{vanzo1}, \cite{louko}, \cite{vanco2}.

In this paper we would like to investigate properties of
topological black holes in the context of the formalism of
approaching quantum gravity developed in \cite{aj1}. This
formalism has its roots in the David Bohm's approach to quantum
mechanics (see e.g., \cite{bohm} and \cite{holland}) and has been
extended to the case of quantum gravity in \cite{jv} and \cite{sq}
(in minisuperspace) and in \cite{jbohm}, \cite{shtanov} (for full
theory.)

The starting point of our investigation will be the Hamiltonian
constraint of the theory, which is the Hamiltonian constraint of
general relativity modified by ``quantum potential'' (for
derivation of this formula, see \cite{aj1}): $$ 0={\cal H}_\bot =
\kappa^2 G_{abcd}p^{ab}p^{cd} +$$ $$ {\cal F}\left(
\frac{27}{16}\frac{\rho^{(5)}{}^2\kappa^2}{\Lambda^2} \sqrt h
+\frac{1}{\kappa^2} \sqrt h R\right.$$ \begin{equation}\left. -
\frac{8}{9} \frac{\Lambda^2}{\kappa^6\rho^{(5)}{}^2} \sqrt h
\left( -\frac38 R^2 + R_{ab}R^{ab}\right)\right), \label{qh}
\end{equation}
\begin{equation}
{\cal F} =
\frac12%
\frac{\sin^2(\phi )}{%
\left\{ \cosh\left(  \frac{3 \rho^{(5)}}{\Lambda} {\cal V} +
\frac{4\Lambda}{3 \kappa^4 \rho^{(5)}}{\cal R}\right)
 + \cos(\phi )
\right\}^{2}},
\end{equation}
where $\kappa$ is the gravitational constant, $G_{abcd}$ is the
Wheeler--De Witt metric, $\rho^{(5)}$ is the renormalization
constant, $\Lambda$ is the bare cosmological and $\phi$ is a
parameter characterizing the solution of Wheeler--De Witt
equation, the following formulas are based on.

\section{Solutions}

We are interested in the static case, where momenta are equal to
zero. In this case one of the dynamical equation of the theory
(corresponding to the $00$ component of Einstein field equations)
is the requirement that the Hamiltonian constraint with $p^{ab}=0$
vanishes, to wit $$
\frac{27}{16}\frac{\rho^{(5)}{}^2\kappa^2}{\Lambda^2} \sqrt h
+\frac{1}{\kappa^2} \sqrt h R$$  \begin{equation}-\frac{8}{9}
\frac{\Lambda^2}{\kappa^6\rho^{(5)} {}^2} \sqrt h \left( -\frac38
R^2 + R_{ab}R^{ab}\right) = 0 \label{G00}
\end{equation}
It is worth observing that the cosmological and renormalization
constants appear only in combination $v_0 =
\frac{\Lambda}{\rho^{(5)}}$ of dimension $[m]^{3}$. Thus the
theory possesses two length scales, the Planck scale $l$ defined
by $\kappa$, and  $v_0^{1/3}$.

For the metric of  (\ref{1}), the solution takes the form
\begin{equation}\label{2}
A=(k+f(r)r^2)^{-1} , \quad k=1,0,-1
\end{equation}
where
\begin{equation}\label{5}
 f(r)= p \pm
\sqrt{\frac34p^2+\frac{\alpha}{r^3}},
\end{equation}
and $p= \frac94\frac{l^4}{v_0^2}$ is a dimensionful parameter,
whose interpretation will be found below. To find the space-time
metrics we make the anzatz $$ ds^2 = -N(r) dt^2 + ds^2_3$$ and
make use of the following variational principle, resulting from
the ADM action in the gauge where the shift vector $N^a=0$, $$I
=-\frac14 \int N{\cal H}_\bot.$$ As a result we find
\begin{equation}\label{6}
 ds^2 = - A(r)^{-1} dt^2 + A(r)dr^2 + d\Omega^2.
\end{equation}

\section{Properties of the metric}

The metric (\ref{6}) should reduce to the metric of the Einstein
topological black hole in the limit $r\rightarrow\infty$. Indeed,
one would expect that the quantum gravity modifications should be
small at large distances. This is indeed the case. For large $r$
we have
\begin{equation}\label{7}
 r^2 f(r) {\rightarrow}
 r^2 p\left(1\pm\frac{\sqrt{3}}{2}\right) \pm \frac{2\alpha}{3pr}.
\end{equation}
For topological black hole we have $$ A(r) = \left(k -
\frac{2M}{r} - \frac{\lambda r^2}{3}\right)^{-1},$$ and thus
asymptotically, our solution corresponds to the topological black
hole in the Anti de Sitter space with cosmological constant
\begin{equation}\label{8}
  \lambda = -
3p\left(1\pm\frac{\sqrt{3}}{2}\right)
\end{equation}
and the mass
\begin{equation}\label{9}
M = \mp \frac{2\alpha}{3p}.
\end{equation}
We see therefore that the parameter $p$ is to be interpreted (up
to the numerical factor) as the physical cosmological constant.

 In what follows we will consider only the case of positive $M$
 and $\alpha$. It can be checked that in the other case (upper
 sign), the space time develops a circular singularity at $r =
 \sqrt[3]{\frac{3p^2}{4|\alpha|}}$. Moreover in this case there is
 no horizon for $k=0,1$. We will investigate such a situation in a
 separate paper.

Let us now turn to horizons of our black holes. They are given as
zeros of the function $A^{-1}(r)$. In the case $k=0$ the position
of horizon can easily be computed explicitly, to wit
\begin{equation}\label{10}
 r_h = \sqrt[3]{\frac{4\alpha}{p^2}}
\end{equation}
The relation $$r_h \sim \alpha^{1/3}$$  holds also in the cases
$k=\pm1$ for large values of $\alpha$. The Killing vector
$\partial/\partial t$ is timelike for $r>r_h$ and spacelike for
$r<r_h$. In particular, the singularity at $r=0$ is spacelike as
in the Schwartzschild black hole.

In the case  $k=+1$  there are two horizons at $r_-$, $r_h$,
corresponding to two real positive roots of the equation $A(r)=0$.
There is only one solution of this equation for
$\alpha=\alpha_{crit}=
     {{\sqrt{\frac{80\,p}{27} +
          \frac{56\,{\sqrt{7}}\,p}{27}}}}$
           and no horizon for smaller
$\alpha$.

In the case $k=-1$ we have one horizon at $r=r_h$ for all values
of $\alpha$. It is clear that $$r_h > r_{crit} = \frac1{\sqrt
p}\,\sqrt{\frac{2}{2-\sqrt{3}}},$$ which is the minimal horizon
radius, corresponding to $\alpha=0$.

Thus the singularity is timelike for $k=1$ and spacelike for
$k=-1$ and is naked for $k=1$ and $\alpha<\alpha_{crit}$.

\section{Thermodynamics}

For metrics of the form (\ref{6}) the standard procedure leads to
the expression for temperature
\begin{equation}\label{11}
 T = \frac{1}{4\pi } \left(\frac{\partial A^{-1}(r)}{\partial
 r}\right)_{r=r_h}.
\end{equation}
Substituting (\ref{2}, \ref{5}) we find
\begin{equation}\label{12}
 T =
\frac{r_h}{4\pi}\left( 2\,p - \frac{{\sqrt{3\,p^2 +
\frac{4\,\alpha}{r_h^3}}}\,
     \left( \alpha + 3\,p^2\,r_h^3 \right) }{4\,\alpha +
     3\,p^2\,r_h^3}\right),
\end{equation}

 and asymptotically for large masses
the relation $T \sim r_h$ i.e., $\alpha \sim T^3$ holds.

Let us now  calculate the entropy. We will proceed exactly as in
\cite{Jurekbh}. The Euclidean action for our solution is free
energy divided by temperature
\begin{equation}
I_E = \frac{M}{T} -S.
\end{equation}
It is well known that $I_E$ consists of bulk integral and the
boundary terms $B$ at infinity and at $r = r_h$ which are fixed by
boundary conditions of the variational problem for the action
$I_E$. Thus we consider $$ I_E = -\int_{r_+}^{\infty} N_0 F' + B,
 $$ where $$F(r) =\frac{1}{4l^2p} \frac{(kA-1)^2}{A^2r} +
\frac{r}{2l^2}\frac{kA-1}{A}
 +\frac{p}{8l^2} r^3.$$ The integral is a linear combination of
constraints ${\cal
H}_\bot(r)$, and thus
\begin{equation}
B =  \frac{M}{T} -S.
\end{equation}
Now it follows that
\begin{equation}
B = \frac{M}{T} + 4 \pi \int\, dr_+ \left(\frac{\partial
F}{\partial A^{-1}(r)}\right)_{r=r_+} - S_0,
\end{equation}
where $S_0$ is a constant to be fixed in a moment. The entropy of
the black hole of outer radius $r_h$ is therefore equal $$
  S = -4
\pi \int\, dr_+ \left(\frac{\partial F}{\partial
A^{-1}(r)}\right)_{r=r_+} + S_0=$$
\begin{equation}\label{13}
=k\frac{2\pi}{9l^2p} \log{r_h} + \frac{\pi}{l^2}r_h^2 + S_0.
\end{equation}

For $k=0$ the temperature vanishes for $r_h=0$ and the logarithmic
term in (\ref{13}) is not present. Therefore we put $S_0=0$ in
this case, and the entropy is purely of the Beckenstein--Hawking
form.

In the case $k=-1$ situation is less clear. Let us observe that
for the minimal radius of the horizon $r_{crit} = \frac1{\sqrt
p}\,\sqrt{\frac{2}{2-\sqrt{3}}}$, the temperature does not vanish.
Nevertheless it seems reasonable to assume that in this case the
temperature $$T(r_{crit}) = \frac{r_{crit} p}{4\pi} \left(2 -
\sqrt3\right)$$ is the minimal temperature and to take $S_0$ so
that
\begin{equation}\label{14}
S^{(-1)} =-\frac{2\pi}{9l^2p}
\log\left(\frac{r_h}{r_{crit}}\right) + \frac{\pi}{l^2}\left(r_h^2
- r_{crit}^2\right),
\end{equation}
as in the case $k=1$, \cite{Jurekbh}. It can be easily checked
that so defined entropy is always positive.

\end{document}